%% file: main.tex
\documentclass{Interspeech2024}
\usepackage{multirow}
\usepackage{cite}
\usepackage{stfloats}
\usepackage{float}
\usepackage{hyperref}
\usepackage{tikz}
\usepackage{graphics}
\usetikzlibrary{positioning}
\usetikzlibrary{shapes.geometric}
\usetikzlibrary{calc}
\usepackage{makecell}

\interspeechcameraready

\title{Codec-ASR: Training Performant Automatic Speech Recognition Systems with Discrete Speech Representations}

\name{Kunal}{Dhawan}
\name{Nithin Rao}{Koluguri}
\name{Ante}{Jukić}
\name{Ryan}{Langman}
\name{Jagadeesh}{Balam}
\name{Boris}{Ginsburg}

\address{NVIDIA Corporation, Santa Clara, CA, USA}
\email{kdhawan@nvidia.com}


\newcommand{\red}[1]{\textcolor{red}{#1}}
\newcommand{\green}[1]{\textcolor{teal}{#1}}

\begin{document}

\maketitle

\begin{abstract}

   Discrete speech representations have garnered recent attention for their efficacy in training transformer-based models for various speech-related tasks such as automatic speech recognition (ASR), translation, speaker verification, and joint speech-text foundational models. In this work, we present a comprehensive analysis on building ASR systems with discrete codes. We investigate different methods for codec training such as quantization schemes and time-domain vs spectral feature encodings. We further explore ASR training techniques aimed at enhancing performance, training efficiency, and noise robustness. Drawing upon our findings, we introduce a codec ASR pipeline that outperforms Encodec at similar bit-rate. Remarkably, it also surpasses the state-of-the-art results achieved by strong self-supervised models on the 143 languages ML-SUPERB benchmark despite being smaller in size and pretrained on significantly less data.

\end{abstract}

\section{Introduction}
\label{sec:intro}
A tremendous amount of progress has been achieved in the area of speech and audio technologies in recent years, in large part due to advances in deep learning and the availability of large-scale datasets~\cite{hinton2012deep, chan2021speechstew, park2022review}.
In particular, transformer-based models led to significant improvements in speech-related tasks such as automatic speech recognition (ASR)~\cite{zhang2020transformer,gulati2020conformer} and joint speech-text modeling~\cite{wang2023neural,wang2023speechx}.

Typically, the input speech signal of an ASR model is represented using a \mbox{mel-spectrogram}, resulting in a continuous representation of the speech signal.
Learnable alternatives have been explored for different applications~\cite{sainath2017multichannel,luo2018tasnet, won2020data, zeghidour2021leaf}.
However, using \mbox{mel-spectrograms} is still a prevalent choice for ASR systems due to their effectiveness~\cite{synnaeve2020end, prabhavalkar2023end}.
Recently, the use of discrete speech representations has garnered attention for their efficacy in training transformer-based models for various speech-related tasks~\cite{wang2023speechx, puvvada2024discrete, chang2023exploration, chang2023exploring} and compatibility with language-modeling architectures~\cite{wang2023neural}.

Discrete speech representations are typically categorized as either acoustic or semantic.
The former capture the acoustic properties of the speech signal, such as pitch, tone, and rhythm. On the other hand, the latter capture the semantic properties of the speech signal, like the meaning and context conveyed by the speech, including words, phrases, and their associations.
Semantic codes are typically obtained by clustering the speech representation at the output of a pre-trained encoder~\cite{hsu2021hubert, chung2021w2v, chen2022wavlm}, or using a codec model~\cite{huang2023repcodec}.
Acoustic codes are typically obtained by compressing and quantizing the speech signal, e.g., using an audio codec, and aim to reconstruct the original signal from a compressed representation.
Several neural audio codecs (NACs) have been proposed recently~\cite{zeghidour2021soundstream, defossez2022high, kumar2023highfidelity, wu2023audiodec, zhang2023speechtokenizer}.
Typically, such codecs have an encoder-quantizer-decoder architecture, where the encoder compresses the input speech signal into a latent representation, quantizer approximates it using a discrete representation, and the decoder reconstructs the original signal from the discrete representation.
Acoustic codes are particularly relevant for multi-task foundational models, which aim to simultaneously understand the content in the input signal and generate high-quality output signals.
While acoustic tokens have been explored in the context of speech and audio synthesis~\cite{wang2023neural, borsos2023soundstorm} and processing~\cite{wang2023speechx}, their use in ASR systems has been relatively underexplored~\cite{puvvada2024discrete}.

To address the above gap, we perform a comprehensive analysis on building ASR systems with discrete codes.
Firstly, we train and evaluate codecs operating in either time or spectral domain with different quantizers.
Secondly, we explore different approaches to improve the ASR system performance, training efficiency and also evaluate approaches for improving their noise robustness. 
Based on our findings, we present a pipeline for noise-robust ASR training with discrete representations generated using a neural audio codec.
Thirdly, to prove the generalizabilty of the proposed NAC+ASR pipeline, we further experiment with the ML-SUPERB dataset~\cite{shi2023ml} consisting of 143 languages. 
The presented results give us a better understanding of the various components of the NAC+ASR pipeline.

We demonstrate that the proposed pipeline based on above learnings is very competitive, outperforming the prevalent Encodec\cite{defossez2022high}-based systems in comparable settings. Our system also achieves a CER of 21\% on the hard ML-SUPERB 1h test set, beating previous state-of-the-art (SOTA) results. The trained NAC\footnote{\url{https://catalog.ngc.nvidia.com/orgs/nvidia/teams/nemo/models/audio_codec_16khz_small}} and ASR models along with accompanying code will be released in the open-source NVIDIA NeMo\footnote{\url{https://github.com/NVIDIA/NeMo}} toolkit.  
\vspace{-5pt}
\section{Speech recognition with audio codecs}
\label{sec:asr_codec}
\input{schemes/codec_architecture}
In this section, we discuss the various components of the proposed ASR pipeline that operates on discrete speech representations.
The block scheme of the complete pipeline is depicted in Figure~\ref{fig:asr_pipeline}.
\vspace{-5pt}
\subsection{Audio codecs}
\vspace{-5pt}
\label{subsec:audio_codec}
Audio codecs capture details of the audio signal using discrete codes at a low bitrate, and are used for speech representation in various tasks, efficient data transmission, and general data compression.
Here we consider two types of NACs, operating either on the time-domain signal or on a spectral domain.
Figure~\ref{fig:codec_architecture} depicts the general architecture of the considered codecs.
\vspace{-6pt}
\subsubsection{Quantization schemes}
\label{subsec:quantization}
\vspace{-5pt}
Residual vector quantization (RVQ) is the common approach used for NAC, e.g., in SoundStream~\cite{zeghidour2021soundstream}, Encodec~\cite{defossez2022high}, and DAC~\cite{kumar2023highfidelity}.
The RVQ uses a series of codebooks with size $D_\text{cb}$, with the current codebook quantizing the residual from the previous quantization step~\cite{zeghidour2021soundstream}.
For each time step, RVQ produces $N_\text{cb}$ codes, corresponding to the number of codebooks.
In this paper, RVQ is configured using $D_\text{enc} = 128$, $D_\text{cb} = 1024$ and $N_\text{cb} = 8$.

Finite scalar quantization (FSQ)~\cite{mentzer2024finite} typically uses a smaller latent dimension $D_\text{enc}$ as compared to RVQ.
Each element of the latent vector is quantized independently into a number level, e.g., to $\{-1,0,1\}$ when using three levels.
As opposed to RVQ, FSQ results in a flat codebook, without a recursive relationship between individual codes.
In this paper, FSQ is configured using $D_\text{enc} = 32$ and $N_\text{cb} = 8$.
For convenience, each $D_\text{enc}/N_\text{cb}$-dimensional subset of the embedding is seen as a separate group quantized with $\left[ 8, 5, 5, 5 \right]$ levels, resulting in $D_\text{cb} = 1000$~\cite{mentzer2024finite}.
\vspace{-6pt}
\subsubsection{Time-domain NAC}
\vspace{-5pt}
\label{subsubsec:time_domain_codec}
Time-domain NAC (TD-NAC) follows the architecture used in previous works~\cite{zeghidour2021soundstream, defossez2022high, wu2023audiodec, kumar2023highfidelity, zhang2023speechtokenizer}.
The encoder consists of a series of convolutional layers with downsampling applied directly on the time-domain signal at sample rate $f_s$, resulting in total downsampling factor $f_\text{down}$.
For each time step, the encoder generates a latent representation of the input signal of dimension $D_\text{enc}$ at rate $f_\text{enc} = f_s / f_\text{down}$, which is quantized to obtain discrete codes.
For reconstructions, discrete codes are dequantized into a latent representation, and a convolutional decoder is used to obtain a time-domain output signal.
Our encoder and decoder configuration is following~\cite{defossez2022high}.
The encoder consists of 1D convolutions followed by residual convolution blocks with downsampling, with LSTM layers for sequence modeling and a final 1D convolution.
The decoder uses a reverse layer ordering with transposed convolutions~\cite{defossez2022high}.
\vspace{-6pt}
\subsubsection{Spectral NAC}
\vspace{-5pt}
\label{subsubsec:spectral_codec}
As opposed to the time-domain NAC, a spectral NAC~\cite{langman2024spectralcodec} applies the encoder on a spectral representation of the input signal obtained using a filterbank as depicted in Figure~\ref{fig:codec_architecture}.
We use an 80-dimensional mel spectrogram obtained from a \mbox{mel-filterbank} and referred to the model as \mbox{Mel-NAC}.

With RVQ we encode the mel-spectrogram with a single residual network consisting of six \mbox{HiFi-GAN V1}~\cite{kong2020hifigan} residual blocks with a hidden dimension of 256 and 1024 residual channels.
With FSQ we divide the mel-spectrogram into 8 groups each containing 10 mel-bands.
Each group is encoded using separate residual encoders with hidden dimension of 128 and 256 residual channels.
The decoder is the \mbox{HiFi-GAN V1} generator with 1024 initial channels.

\begin{figure}[b] 
  \centering
  \resizebox{\columnwidth}{!}{
  \includegraphics{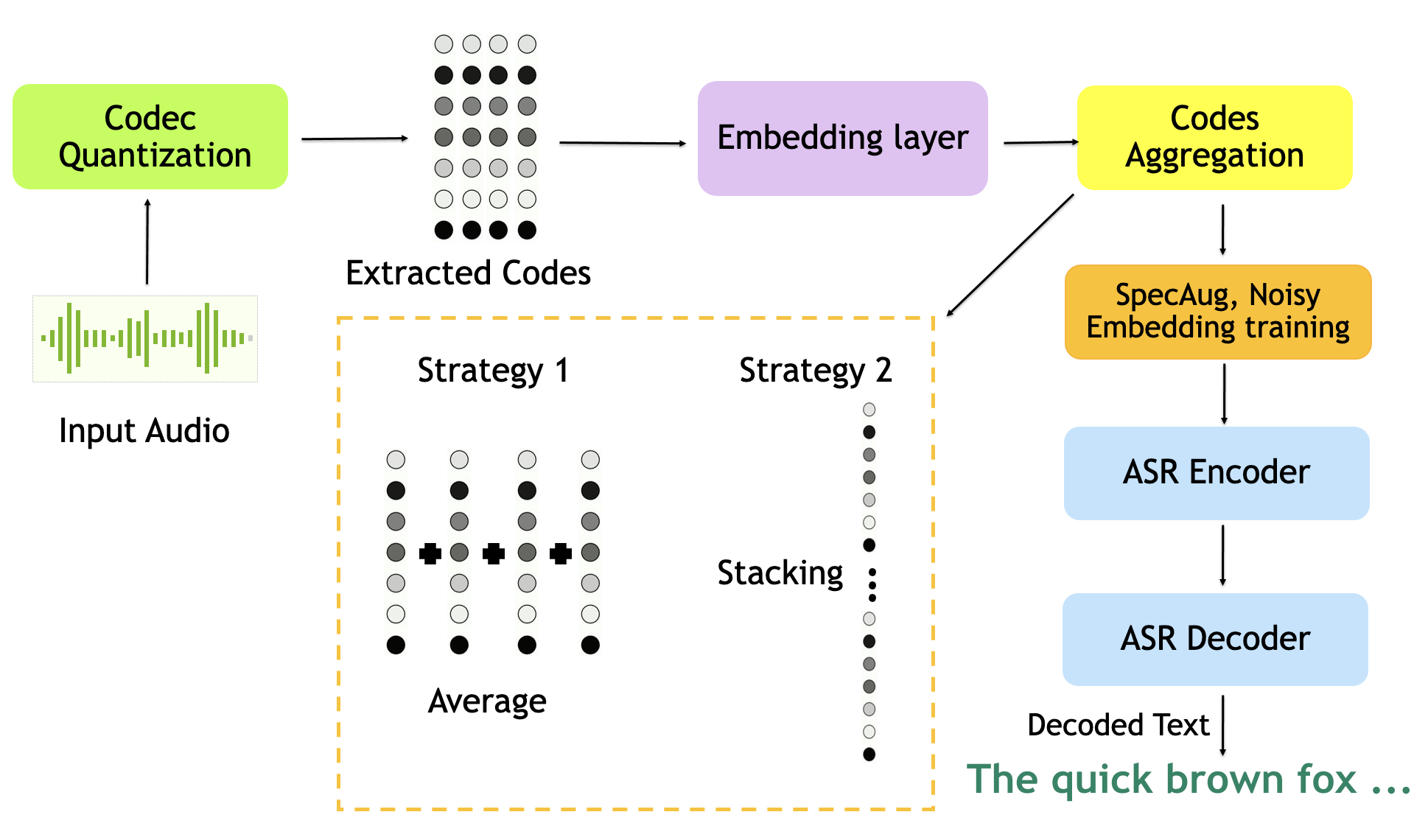} 
  }
  \caption{The ASR with discrete codes pipeline.}
 \label{fig:asr_pipeline}
\end{figure}

\vspace{-5pt}
\subsection{Speech recognition pipeline}
\vspace{-5pt}
\subsubsection{Embedding layer and codebook initialization}
\label{subsubsec:embedding_layer}
\vspace{-5pt}
The initial stage of the pipeline involves the mapping of codes to embeddings, which are subsequently forwarded to the ASR encoder for model training. Here we employ a standard neural embedding layer which maps the output of each codebook to a fixed dimensional embedding of size $D_\text{emb}$. The parameters of this neural embedding model are iteratively optimized during the end-to-end ASR system training. We can either initialize the weights of the embedding model randomly or use the learnt codebooks from the trained NAC model to provide a better starting point. We refer to the latter approach as codebook initialization of the embedding layer in the rest of the paper. 

\vspace{-8pt}
\subsubsection{Code aggregation strategies}
\label{subsubsec:code_agg_strategy}
\vspace{-5pt}
As discussed in Section~\ref{subsec:audio_codec}, most NACs employ multiple codebooks to obtain reliable compressed discrete representation of the input signal. Consequently, this results in the presence of multiple codes per time step corresponding to each codebook. It becomes imperative to aggregate across codebooks for each timestep to enable their integration into standard ASR encoder-decoder architectures. This aggregation process can be executed through two distinct schemes, as illustrated in Figure~\ref{fig:asr_pipeline}: stacking and averaging. In the stacking (stack) aggregation approach, embeddings from different codebooks are stacked atop one another, yielding an embedding size of $N_{cb} \times D_\text{emb}$. Conversely, the averaging (avg) aggregation approach entails the computation of the average of embeddings from different codebooks at each timestamp, resulting in an embedding size of $D_\text{emb}$. In this paper, the default codes aggregation strategy is averaging, unless otherwise specified.

\vspace{-8pt}
\subsubsection{Spectrogram augmentation}
\vspace{-5pt}
The technique of spectrogram augmentation (SpecAug) serves as a method for augmenting audio data, as introduced in~\cite{park2019specaugment}. This methodology transforms the augmentation task for audio signals into one resembling image augmentation by operating on the audio spectrogram. Though in this work we are training the ASR systems on discrete codes, we evaluate the impact of SpecAug on the ASR pipeline. 

\vspace{-5pt}
\subsubsection{Noisy embedding training}
\label{subsubsec:noisy_emb_training}
\vspace{-5pt}
Advancements in large language model (LLM) research has shown that model fine-tuning process can be improved by the simple augmentation technique of adding noise to the embedding vectors during training~\cite{jain2023neftune}. We evaluate the efficacy of this method by adding scaled uniform noise (parameterized by $\alpha$ as introduced in~\cite{jain2023neftune}) to the output of the embedding layer (Section~\ref{subsubsec:embedding_layer}) during the training phase.

\vspace{-5pt}
\section{Experimental setup}
\vspace{-2pt}

\begin{table}[t]
    \caption{Configurations of the considered NACs.}
    \label{tab:td-nac-s-nac}
    \vspace{-0.8em}
    \centering
    \resizebox{0.82\columnwidth}{!}{%
    \begin{tabular}{ccccc}
        \toprule
        Codec   & Quantizer & Parameters / $10^6$ & $D_\text{enc}$ & $f_\text{enc}$ / Hz \\ \midrule
        TD-NAC  & RVQ       & 13.8                & 128            & 80 \\ \midrule
        TD-NAC  & FVQ       & 13.1                & 32             & 80 \\ \midrule
        Mel-NAC & RVQ       & 105                 & 128            & 62.5 \\ \midrule
        Mel-NAC & FVQ       & 104                 & 32             & 62.5 \\
        \bottomrule
    \end{tabular}
    }
\end{table}
\subsection{NAC model training}
\label{subsec:exp_NAC}
\vspace{-5pt}
Both \mbox{TD-NAC} and \mbox{Mel-NAC} are trained on the \mbox{Libri-Light} dataset~\cite{kahn2020libri} with sample rate 16\,kHz.
\mbox{TD-NAC} models use an encoder with downsampling rates of $\left\lbrace 2,4,5,5 \right\rbrace$, resulting in $f_\text{enc} = 80$\,Hz.
Both RVQ and FSQ quantizers use $N_\text{cb}=8$ codebooks with $D_\text{cb} \approx 2^{10}$, resulting in a bitrate of 6.4\,kbps.
The \mbox{TD-NAC} decoder upsamples in the reverse order of $\left\lbrace 5,5,4,2 \right\rbrace$ to obtain the reconstructed audio signal.
The model is trained on examples with one second of audio.
\mbox{Mel-NAC} models use mel-filterbank with a frame length of 1024 samples and frame shift of 256 samples, resulting in $f_\text{enc} = 62.5$\,Hz.
Using the same quantizer setup as for \mbox{TD-NAC}, this results in a bitrate of 5\,kbps.
The \mbox{Mel-NAC} decoder upsamples at rates of $\left\lbrace 8,4,4,2 \right\rbrace$ to obtain the reconstructed audio signal.
The model is trained on examples with 0.512 seconds of audio.
All NAC models are trained end-to-end using time-domain loss, discriminative loss, and frequency-domain loss, similar to~\cite{defossez2022high} with equal weights for frequency and discriminative loss and $0.1$ weight for time-domain loss.
Model sizes depending on the corresponding quantizer are provided in Table~\ref{tab:td-nac-s-nac}.
The models are trained on eight NVIDIA V100 GPUs for 130k steps with the AdamW optimizer with a learning rate of $10^{-4}$.
A StepLR scheduler with a step size of 1 and gamma of 0.999996 is employed for learning rate decay.
\vspace{-4pt}
\subsection{ASR model training}
\label{subsec:exp_ASR}
\vspace{-5pt}

The ASR models presented in the paper adopt the FastConformer Transducer large architecture~\cite{rekesh2023fast} with 114\,M parameters. The encoder consists of 17 layers, with a model dimension of 512. We used 256 channels in sub-sampling module and a kernel size of 9 in convolution module. A single layer RNN-T with hidden dimension of 640 is used for decoder. We maintain the embedding layer output dimension $D_\text{emb}$ (Section~\ref{subsubsec:embedding_layer}) at 128 and set $\alpha$ (Section~\ref{subsubsec:noisy_emb_training}) to 5 across all experiments to ensure equitable comparison. The ASR models are trained on the LibriSpeech corpus~\cite{panayotov2015librispeech}, encompassing 960 hours of English speech data. Evaluation of ASR model performance is conducted using the standard 'clean' and 'other' sets of dev and test partitions from the LibriSpeech dataset. We use a Sentencepiece Byte Pair Encoding (BPE)~\cite{kudo2018sentencepiece} tokenizer with a vocabulary size of 1024, trained on the text data from the LibriSpeech training set. All ASR models have been trained for 100k updates on two nodes with eight NVIDIA A100 80GB GPUs using a batch size of 32 on each GPU.
We use AdamW with a peak learning rate of $2 \cdot 10^{-3}$, 15k warmup steps with Cosine annealing, minimum learning rate of $10^{-6}$ and weight decay of $10^{-3}$. 
\vspace{-4pt}
\subsection{Experiments and ablations}
\label{subsec:exp}
\vspace{-5pt}
The experiments are designed to study and understand four major components of the pipeline: (\textit{i}) role of the NAC type, i.e., \mbox{TD-NAC} vs \mbox{Mel-NAC}, (\textit{ii}) role of quantizers in NAC, i.e., RVQ vs FSQ, (\textit{iii}) effect of code aggregation strategies, (\textit{iv}) performance improvements of codec ASR systems with pipeline optimizations.
We also setup strong baselines in the form of the traditional Mel-Spectrogram features as well as the widely used Encodec audio codec~\cite{defossez2022high}.
All other components like $D_\text{emb}$, ASR model size, ASR training data, and tokenizer have been kept constant to facilitate an unbiased study towards the role played by the above highlighted four components. Word error rate (WER) metric is used to evaluate the performance of the ASR models.
\vspace{-5pt}
\section{Results and discussion}
\label{sec:results}
\begin{table}[t]
    \caption{ASR improvement on LibriSpeech eval sets contributed by the various components of the presented ASR pipeline.}
    \label{tab:asr_improvement}
    \vspace{-0.8em}
    \resizebox{\columnwidth}{!}{%
        \begin{tabular}{lcccc}
            \toprule
                       & \multicolumn{4}{c}{\textbf{WER / \% $\downarrow$}} \\ \cmidrule(lr){2-5} 
            Codec      & \textbf{dev-clean} & \textbf{dev-other} & \textbf{test-clean} & \textbf{test-other} \\ \midrule
            TD-NAC-RVQ & 17.58              & 38.77              & 17.18               & 41.55               \\ \midrule
            \makecell[l]{+~codebook\\\hspace{0.75em}initialization} & \makecell[c]{3.87\\(\green{-13.71})} & \makecell[c]{12.17\\(\green{-26.6})} & \makecell[c]{3.84\\(\green{-13.34})} & \makecell[c]{12.28\\(\green{-29.27})} \\ \midrule
            \makecell[l]{\hspace{0.5em}+~spectrogram\\\hspace{1.25em}augmentation} & \makecell[c]{2.21\\(\green{-1.66})} & \makecell[c]{5.83\\(\green{-6.34})} & \makecell[c]{2.36\\(\green{-1.48})} & \makecell[c]{5.84\\(\green{-6.44})} \\ \midrule
            \makecell[l]{\hspace{1.0em}+~noisy embedding\\\hspace{1.75em}training} & \makecell[c]{2.19\\(\green{-0.02})} & \makecell[c]{5.72\\(\green{-0.11})} & \makecell[c]{2.4\\(\red{+0.04})} & \makecell[c]{5.76\\(\green{-0.08})} \\
            \bottomrule
        \end{tabular}
    }
\end{table}

\begin{table*}[t]
\caption{ASR performance on LibriSpeech evaluation sets for the considered pipeline configurations.}
\label{tab:asr_main}
\vspace{-0.9em}
\centering
\resizebox{0.95\textwidth}{!}{%
\begin{tabular}{cccccccccccc}
    \toprule
    \multirow{2}{*}{\shortstack{Input feature}} & \multirow{2}{*}{Quantizer} & \multirow{2}{*}{$f_s$ / kHz} & \multirow{2}{*}{Bitrate / kbps} & \multirow{2}{*}{\shortstack{Code\\aggregation}} & \multirow{2}{*}{$N_\text{cb}$} & \multirow{2}{*}{$D_\text{cb}$} & \multirow{2}{*}{$D_\text{enc}$} & \multicolumn{4}{c}{WER / \% $\downarrow$} \\ \cmidrule(lr){9-12}
                    &     &    &     &       &    &      &     & dev clean & dev-other & test clean & test other \\ \midrule
    mel-spectrogram & --  & -- & --  & --    & -- & --   & --  & 2.12 &  4.88 & 2.27 &  5.03 \\ \midrule
    EnCodec         & RVQ & 24 & 24  & avg   & 32 & 1024 & 128 & 2.16 &  5.68 & 2.3  &  5.47 \\ 
    EnCodec         & RVQ & 24 & 12  & avg   & 16 & 1024 & 128 & 2.26 &  5.77 & 2.45 &  5.8  \\ 
    EnCodec         & RVQ & 24 & 6   & avg   & 8  & 1024 & 128 & 2.23 &  6.02 & 2.35 &  5.96 \\ 
    EnCodec         & RVQ & 24 & 3   & avg   & 4  & 1024 & 128 & 2.44 &  7.13 & 2.6  &  7.13 \\ \midrule
    TD-NAC          & RVQ & 16 & 6.4 & stack & 8  & 1024 & 128 & 3.12 & 10.17 & 3.38 & 10.17 \\ 
    TD-NAC          & RVQ & 16 & 6.4 & avg   & 8  & 1024 & 128 & 2.19 &  \textbf{5.72} & \textbf{2.40}  &  \textbf{5.76} \\ 
    TD-NAC          & FSQ & 16 & 6.4 & stack & 8  & 1000 & 32  & \textbf{2.18} &  6.08 & 2.42 &  5.92 \\ \midrule
    Mel-NAC         & RVQ & 16 & 5   & avg   & 8  & 1024 & 128 & 2.23 &  5.92 & 2.40 &  5.80 \\ 
    Mel-NAC         & FSQ & 16 & 5   & stack & 8  & 1000 & 32  & 2.33 &  6.18 & 2.45 &  6.09 \\
    \bottomrule
\end{tabular}

}
\end{table*}
\vspace{-3pt}
\subsection{Codebook initialization, spectrogram augmentation and noisy embedding training}
\label{subsec:results_init}
\vspace{-5pt}
To investigate these components' effects, we first train a TD-NAC model with RVQ following the specifications outlined in Section~\ref{subsec:exp_NAC}. Utilizing features from this audio codec as input, we establish our baseline ASR pipeline, employing parameters detailed in Section~\ref{subsec:exp_ASR}, yielding the baseline performance noted in the first row of Table~\ref{tab:asr_improvement}.
Subsequently, we adapt the ASR pipeline to initialize the embedding layer with codebooks learned from the trained NAC (Section~\ref{subsubsec:embedding_layer}), maintaining other pipeline components unchanged. With this setup, we train another ASR system and report it's performance in the second row of Table~\ref{tab:asr_improvement}.  Likewise, we progressively integrate spectrogram augmentation and noisy embedding training into the pipeline. Notably, codebook initialization of the embedding layer significantly enhances the ASR model's performance, with more than 10\% absolute WER improvement across all the evaluation sets. Spectrogram Augmentation aids in enhancing the model's noise robustness, as reflected by more than 6\% absolute WER improvement on the noisy 'other' sets. Noisy embedding training is able to even further improve this noise robustness of the model. Consequently, for all subsequent experiments, we incorporate all three components - codebook initialization, spectrogram augmentation, and noisy embedding training - into the training pipeline.
\vspace{-4pt}
\subsection{Code aggregation strategy}
\label{res:code_agg}
\vspace{-5pt}
To assess the influence of the code aggregation strategy on the ASR+NAC model pipeline, we build up on the baseline setting as motivated in Section~\ref{subsec:results_init}: TD-NAC model with RVQ, FastConformer-RNNT ASR model with embedding layer initialized with the learnt codebooks, SpecAug, and noisy embedding training. Two models are trained: one utilizing stacking for aggregating code embeddings and the other employing averaging (refer to Section~\ref{subsubsec:code_agg_strategy}). The performance of these models are reported in rows $6$ and $7$ of Table~\ref{tab:asr_main}. Notably, the averaging strategy yields significantly superior WER performance compared to stacking. It's worth noting that the embedding dimension $D_\text{emb}$ (as discussed in Section~\ref{subsubsec:embedding_layer}) remained fixed at 128 for both runs and the results might change with an increase in the embedding dimension. However, to ensure a fair comparison and assess the optimal configuration within the described setup, the embedding dimension was kept constant.  

Despite the noted performance, stack remains the preferred aggregation scheme for all our NAC-FSQ systems. This choice is informed by the realization that different FSQ codebooks quantize distinct segments of the encoder output, whereas the RVQ codebooks encode residuals of the same vector.

\vspace{-5pt}
\subsection{Neural audio codec type}
\vspace{-5pt}

We proceed to examine and compare TD-NAC with Mel-NAC, assessing their influence on downstream ASR tasks. Owing to the distinct down-sampling structures and rates outlined in Section~\ref{subsec:audio_codec}, the compared TD-NAC operates at a bit-rate of 6.4\,kbps, whereas Mel-NAC operates at 5\,kbps.
The remainder of the ASR pipeline remains constant, incorporating insights from Section~\ref{res:code_agg}, and we compare both RVQ and FSQ versions of the codecs. The results of these ablations are presented in the last four rows of Table~\ref{tab:asr_main}. Notably, TD-NAC demonstrates slightly better performance compared to Mel-NAC across all considered ASR eval sets. This finding is intriguing, given that Mel-NAC outperforms TD-NAC for TTS tasks~\cite{langman2024spectralcodec}.
Hence, the selection of the NAC should consider the downstream task.

Furthermore, we observe that the presented TD-NAC with RVQ and only 8 codebooks outperforms Encodec with 4, 8, and even 16 codebooks, while maintaining all other parameters such as codebook size and ASR system parameter counts constant. The performance of the TD-NAC system with a bit-rate of only 6.4\,kbps closely matches that of Encodec with 24\,kbps (utilizing all 32 codebooks). We have open-sourced the weights (\textit{audio\_ codec\_16khz\_small}) and code\footnote{\url{https://github.com/NVIDIA/NeMo/blob/main/tutorials/tts/Audio_Codec_Training.ipynb}} for this codec model so that it can be utilized by and be built upon by the community.
\vspace{-5pt}
\subsection{Quantization schemes}
\vspace{-5pt}
Finally, we study the effect of quantization schemes on downstream ASR performance.  Analysis of the last four rows of Table~\ref{tab:asr_main} reveals that FSQ detrimentally affects ASR performance, particularly on the noisy 'other' sets. We hypothesize this happens because of  the fixed finite level encoding scheme utilized by FSQ, which poses challenges in modeling noisy data. 

\vspace{-4pt}
\section{Multilingual extension}
\vspace{-5pt}
To demonstrate the generalization ability of the presented NAC+ASR pipeline, we performed a study using additional languages and broader corpora.
To this end, we participated in the ASR track of the Interspeech 2024 Speech Processing Using Discrete Speech Unit Challenge~\cite{dsu_challenge_2024} that uses the ML-SUPERB~\cite{shi2023ml} dataset comprising of 143 languages.

\vspace{-6pt}
\subsection{Model and data description}
\vspace{-5pt}
Our pipeline uses TD-NAC model with RVQ, as detailed in Section~\ref{subsec:exp_NAC}, that obtained the best performance in the experiments summarized in Table~\ref{tab:asr_main}.
The NAC model was not retrained and we use the same setup as in Section~\ref{subsec:exp_NAC}.
For ASR we use the FastConformer-RNNT model described in Section~\ref{subsec:exp_ASR} along with avg code aggregation strategy, codebook initialization of the embedding layer, SpecAug and noisy embedding training, based on Section~\ref{sec:results}.
As per the challenge requirements, the ASR model is trained on the LibriSpeech-clean-100 subset (100 hrs) along with the \mbox{ML-SUPERB} 1h set (222 hrs) which contains data from 143 languages.
The combined data has 6280 unique characters.
\vspace{-6pt}
\subsection{Results}
\vspace{-5pt}
\begin{table}[]
    \caption{CER on the ML-SUPERB 1h test set.}
    \label{tab:challenge}
    \vspace{-0.2cm}
    \centering
    \resizebox{0.7\columnwidth}{!}{%
    \begin{tabular}{ccc}
        \toprule
        System & Challenge baseline & Our system \\ \midrule
        CER & 72.6 & 21.0 \\
        \bottomrule
    \end{tabular}
    }
\end{table}
We compare the performance of our NAC+ASR pipeline with the baseline system~\cite{dsu_challenge_2024} on the \mbox{ML-SUPERB} 1h test set which consists of 45 hours of unseen speech.
Table~\ref{tab:challenge} presents the Character Error Rate (CER) metric for both systems.
It can be observed that our system with 21\% CER significantly outperforms the challenge baseline.
Moreover, our system surpasses the SOTA performance achieved by the XLSR-128 model, which reported a CER of 22\%~\cite{shi2023ml}, despite being smaller in size and pretrained on significantly less data.
This competitive CER underscores the effectiveness of the proposed NAC+ASR pipeline not only in monolingual scenarios (cf. Table~\ref{tab:asr_main}) but also in multilingual settings encompassing over 100 languages.
\vspace{-4pt}
\section{Conclusion}
\vspace{-5pt}

In this work, we presented a speech recognition pipeline working on discrete codes from an audio codec and performed a study of different components of the system.
We trained neural audio codecs with different quantizers and found that time-domain codec with RVQ resulted in the best performance on the considered data.
We investigated ASR pipeline optimizations and found that optimal code aggregation and codebook initialization resulted in large performance improvements.
Furthermore, we found that SpecAug and noisy embedding training in our pipeline lead to improved performance in clean conditions and superior robustness in noisy conditions.
Our best result outperforms EnCodec-based model at a comparable bit-rate.
Finally, we studied the performance on a large multi-lingual dataset.
The proposed model beats the SOTA performance of strong self-supervised models like XLSR-128 on the 143-language ML-SUPERB benchmark despite being smaller and trained on significantly less data.
All the trained NAC and ASR models along with accompanying code will be released in the NeMo toolkit~\cite{nvidia_nemo_toolkit}.

\cleardoublepage
\bibliographystyle{IEEEtran}
\bibliography{main}

\end{document}

%% file: schemes/codec_architecture.tex
\begin{figure}
    \centering
    \scriptsize
    \resizebox{0.75\columnwidth}{!}{
    \begin{tikzpicture}
      \node[draw,
        rectangle,
        dashed,
        minimum width=2.5cm,
        minimum height=0.4cm,
        fill={rgb:red,118;green,185;blue,0},
        fill opacity=0.1,
        text opacity=1.0,
        align=center
      ] (filterbank) {Filterbank};
      \node[fill=white, left=-1.2cm of filterbank.east] (filterbank_label) {Optional} ;
      \node[] at ($(filterbank.west) + (-1.5,0.5)$) (input) {time-domain input} ;
      \coordinate (input_connect) at ($(filterbank) + (0,0.5)$) ;
      \draw[-latex] (input) -- (input_connect) -- (filterbank) ;
      \node[draw,
        rectangle,
        trapezium, trapezium angle=-60,
        inner xsep=0pt,outer sep=0pt, text width=0.8cm,
        minimum width=2.5cm,
        minimum height=0.4cm,
        fill={rgb:red,118;green,185;blue,0},
        fill opacity=0.2,
        text opacity=1.0,
        below=0.3cm of filterbank,
        align=center
      ] (encoder) {Encoder};
      \draw[-latex] (filterbank) -- (encoder) ;
      \node[below=0.1cm of encoder] (before_quantize) {};
      \draw[-] (encoder) -- (before_quantize.center);
      \node[draw,
        rectangle,
        minimum width=1.65cm,
        minimum height=0.4cm,
        fill={rgb:red,118;green,185;blue,0},
        fill opacity=0.2,
        text opacity=1.0,
        below=0.1cm of before_quantize,
        align=center
      ] (quantize) {quantize};
      \draw[-latex] (before_quantize.center) -- (quantize);
      \node[draw,
        rectangle,
        dotted,
        minimum height=0.6cm,
        minimum width=3.8cm,
        fill=white,
        below=0.4cm of quantize,
        align=center
      ] (tokens) {};
      \draw[-latex] (quantize) -- (tokens);
      \node[fill=white, left=0.1cm of tokens.west] (tokens_label) {codes} ;
      \node[draw,
        circle,
        minimum height=0.3cm,
        fill=black!60,
        align=center
      ] at (tokens) (token_center) {};
      \node[draw,
        circle,
        minimum height=0.3cm,
        fill=black!20,
        left=0.2cm of token_center,
        align=center
      ] (token_left_1) {};
      \node[draw,
        circle,
        minimum height=0.3cm,
        fill=black!5,
        left=0.2cm of token_left_1,
        align=center
      ] (token_left_2) {};
      \node[draw,
        circle,
        minimum height=0.3cm,
        fill=black!98,
        left=0.2cm of token_left_2,
        align=center
      ] (token_left_3) {};
      \node[draw,
        circle,
        minimum height=0.3cm,
        fill=black!50,
        right=0.2cm of token_center,
        align=center
      ] (token_right_1) {};
      \node[draw,
        circle,
        minimum height=0.3cm,
        fill=black!90,
        right=0.2cm of token_right_1,
        align=center
      ] (token_right_2) {};
      \node[draw,
        circle,
        minimum height=0.3cm,
        fill=black!10,
        right=0.2cm of token_right_2,
        align=center
      ] (token_right_3) {};
      \node[draw,
        rectangle,
        minimum width=1.65cm,
        minimum height=0.4cm,
        fill={rgb:red,118;green,185;blue,0},
        fill opacity=0.2,
        text opacity=1.0,
        below=0.4cm of tokens,
        align=center
      ] (dequantize) {dequantize};
      \draw[-latex] (tokens) -- (dequantize) ;
      \node[below=0.1cm of dequantize] (before_decoder) {};
      \draw[-] (dequantize) -- (before_decoder.center);
      \node[draw,
        trapezium, trapezium angle=60,
        inner xsep=0pt,outer sep=0pt, text width=0.8cm,
        minimum width=2.5cm,
        minimum height=0.4cm,
        fill={rgb:red,118;green,185;blue,0},
        fill opacity=0.2,
        text opacity=1.0,
        below=0.1cm of before_decoder,
        align=center
      ] (decoder) {Decoder};
      \draw[-latex] (before_decoder.center) -- (decoder) ;
      \node[] at ($(decoder.east) + (1.5,-0.5)$) (output) {time-domain output} ;
      \coordinate (output_connect) at ($(decoder) + (0,-0.5)$) ;
      \draw[-latex] (decoder) -- (output_connect) -- (output) ;
      \node[draw=gray,
        dashed,
        rectangle,
        minimum height=2.6cm,
        minimum width=6.0cm,
        fill={rgb:red,118;green,185;blue,0},
        fill opacity=0.02,
        text opacity=1.0,
        anchor=center
      ] at (tokens) (quantizer) {};
      \node[fill=white!50,
        fill opacity=0.75,
        text opacity=1.0,
        right=-0.5cm of quantizer,
        align=center
      ] (quantizer_label) {Quantizer} ;
    \end{tikzpicture}
    }
    \caption{Architecture of the considered neural audio codecs.}
    \label{fig:codec_architecture}
  \end{figure}